\documentclass{article}
\usepackage{siunitx} 
\usepackage{cite}

\usepackage[pdftex]{graphicx}
\usepackage{subcaption}
\DeclareGraphicsExtensions{.pdf,.jpg,.jpeg,.png}
\usepackage{amsmath}
\usepackage{amssymb}
\usepackage{amsfonts}
\usepackage{algorithmic}
\usepackage{url}

\begin{document}
%
\title{On a Dissimilarity Metric for Analyzing Body Synergistic Coordination in 
Non-Periodic Motion
}
%
%
%

\author{Shunpei Fujii \and
        Kanta Tachibana 
\thanks{S.~Fujii and K.~Tachibana were with the Department
of System Mathematics Science, School of Informatics, Kogakuin University, Japan e-mail:jx21054@g.kogakuin.jp, kanta@cc.kogakuin.ac.jp.}}

\maketitle

\begin{abstract}
This study proposes a novel metric to quantitatively evaluate body synergistic coordination, explicitly addressing dynamic interactions between pairs of body segments in baseball pitching motions. 
Conventional methods typically compare motion trajectories using individual joint coordinates or velocities independently, employing techniques like Dynamic Time Warping (DTW) that inherently apply temporal alignment even when such correction may distort meaningful rhythm-based differences. 
In contrast, our approach models the coordination dynamics as Linear Time-Invariant (LTI) systems, leveraging convolution operations between pairs of time series data to capture the gain and phase-lag inherent in genuine coordination dynamics. 
Empirical validation demonstrates the robustness of the proposed metric to variations in camera angles and scaling, providing superior discriminative capability compared to DTW and deep learning-based methods.
\end{abstract}


%

\section{Introduction}

Analyzing and quantifying body coordination from motion data is essential in various fields, such as sports analytics, rehabilitation, and robotics. 
Traditional methods predominantly focus on evaluating the similarity between entire motion trajectories using individual time series signals (e.g., joint coordinates or velocities) independently. 
Notably, Dynamic Time Warping (DTW) has been widely utilized for time-series comparisons due to its ability to align sequences temporally, effectively addressing differences in movement speed or duration~\cite{Ionescu2014, Shu2022, Martinez2017}. 
However, DTW inherently applies temporal warping even when such compensation is not necessary, for instance in comparing pitches with intentionally rhythmic differences (paused versus quick pitches), potentially distorting genuine coordination dynamics.

Recent studies employing machine learning and deep learning approaches further underscore this limitation, as they typically derive similarity metrics directly from spatio-temporal joint data~\cite{Wang2019, Wang2024, Chopin2021}. 
Methods integrating neural networks, such as Siamese architectures or deep metric learning~\cite{Chen2021, zhang2024human}, primarily embed entire motion sequences into a latent space to quantify similarities but still fundamentally treat individual signals separately, neglecting explicit inter-segment coordination mechanisms.

To overcome these limitations, we propose a novel dissimilarity metric to quantitatively evaluate synergistic coordination between pairs of body segments. 
Instead of directly comparing individual joint trajectories independently, our method evaluates the coordination between pairs of body segments by leveraging their dynamical interactions through convolution operations. 
This approach uniquely models the relationship between input-output pairs as Linear Time-Invariant (LTI) systems, explicitly capturing gain and phase-lag information inherent in the body's synergistic coordination.

Our proposed metric addresses two significant gaps in existing literature:
\begin{enumerate}
    \item It avoids unnecessary temporal alignment corrections inherent in DTW, allowing accurate comparisons of movements with intentionally distinct rhythms.
    \item It explicitly quantifies the dynamic interactions between specific pairs of body segments (such as the ankle and wrist in baseball pitching), enabling precise differentiation between movements performed by the same individual and those performed by different individuals, independent of variations in camera angle or image scale.
\end{enumerate}
Furthermore, we demonstrate the robustness of our metric against variations in camera angle and scaling differences in 2D video-based analysis, challenges commonly encountered in practical scenarios~\cite{Cao2017, Kanno2022, Li2021, Jafarzadeh2021, Osawa2022}.

The remainder of this paper is structured as follows.
Section 2 presents the theoretical formulation of our proposed method. 
Section 3 describes the experimental setup. 
Section 4 reports experimental results.
And Sections 5 and 6 discuss the implications of our findings and conclude the paper, respectively.

\section{Theoretical Background}
To investigate coordination in aperiodic whole-body movements, we first compute the time series data (in this study, the change in coordinates on the screen) representing the movements of two locations near the end of the body (in this study, the wrist of the throwing arm and the ankle of the opposite side) from the video. 
We then Z-transform these time series as inputs and outputs of the LTI system and evaluate the transfer functions to define a quantitative measure of whole-body coordination, specifically a dissimilarity measure derived from these transfer functions. 
The frame rate is assumed to be the same for both videos. 
The number of frames from the start of a movement to the end of the movement is generally different between the two videos. 
In this study, we focus on the speed of the ankle and wrist, respectively, and let $a[n], b[n], n=0,1,\dots,N-1$ be the time series data in the first video, and  $x[m], y[m], m=0,1,\dots,M-1$ be the time series data in the second video. 
Based on the transfer function concept, Z-transform each of them. 
$$\mathcal{Z}: \{a[n],n\in[0,N-1)\} \mapsto A(z)=\sum_{n=0}^{N-1} a[n]z^{-n}$$
Similarly,
$$B(z)=\sum_{n=0}^{N-1} b[n]z^{-n},\hspace{1.5mm} 
X(z)=\sum_{m=0}^{M-1} x[m]z^{-m},\hspace{1.5mm}
Y(z)=\sum_{m=0}^{M-1} y[m]z^{-m}$$
The transfer function from ankle speed to wrist speed in the first video is $$G_1(z)=\frac{B(z)}{A(z)}$$. 
Similarly, the transfer function for the second video is $G_2(z)=Y(z)/X(z)$. 
If the body coordination during the two movements is similar, then the transfer functions $G_1(z)$ and $G_2(z)$ are similar, and $A(z)Y(z)$ and $X(z)B(z)$ are similar. 

As 
\begin{eqnarray}
  A(z)Y(z) &=& \left(\dots+a[n]z^{-n}+\cdots\right) 
  \left(\dots+y[m]z^{-m}+\cdots\right)\\
  &=& \sum_{k=0}^{M+N-2}(a*y)[k]z^{-k}\label{eq:AYproduct},
\end{eqnarray}
the coefficient of each term in the polynomial $A(z)Y(z)$ is equal to the convolution. 
In~(\ref{eq:AYproduct}), setting $a[n]=y[m]=0\hspace{1mm} \forall n\notin[0,N), m\notin[0,M)$,
\begin{equation}
  (a*y)[k] = \sum_{\ell=0}^k a[\ell]y[k-\ell].
\end{equation}
Similarly, setting $x[m]=b[n]=0\hspace{1mm}\forall m\notin[0,M), n\notin[0,N)$, the convolution $(x*b)[k]=\sum_{\ell=0}^k x[\ell]b[k-\ell]$ is defined and 
\begin{equation}
  X(z)B(z)=\sum_{k=0}^{M+N-2}(x*b)[k] z^{-k}
\end{equation}

Let the coefficients of the terms $A(z)Y(z)$ and $X(z)B(z)$ be $(M+N-1)$-dimensional vectors $u=(a*y),v=(x*b)$, respectively.
And we propose a dissimilarity measure, 
\begin{equation}
  \mathrm{Dis}((a,b),(x,y))=\frac{||u-v||^2}{||u||||v||}.
\end{equation}
This dissimilarity measure can evaluate the dissimilarity of body coordination even when the distance from the camera to the subject is different and the scale is different between the pairs $(a,b)$ and $(x,y)$, the time series data obtained from the video.

\section{Method and Experiments}
\subsection{Video-Based Body Coordination Analysis}
We analyzed six publicly available pitching videos from three Major League Baseball (MLB) pitchers, with two videos per pitcher. 
Joint coordinates were estimated using OpenPose. 
The start of the pitching motion is defined as the frame in which the heel of the outstretched foot leaves the ground and the Y (vertically downward) coordinates of the ankle begin to decrease, and the end of the pitching motion is defined as the frame immediately preceding ball release. 

We focus on the ankle (left ankle for the right pitcher and right ankle for the left pitcher) and wrist (right wrist for the right pitcher and left wrist for the left pitcher) and calculate the difference $\Delta x[n]$ in the X coordinate and $\Delta y[n]$ in the Y coordinate between frame $n-1$ and frame $n$. 
And we calculate speed for $n$-th frame $s[n]=\sqrt{(\Delta x[n])^2+(\Delta y[n])^2}$ [pixels/frame].
We calculate the ankle and wrist speeds $s_a[n], s_w[n]$. 
Then, the $(s_a,s_w)$ dissimilarity is calculated as described in the theoretical background for each pair of videos.

\begin{figure}[htbp]
    \centering
    \includegraphics[width=\linewidth]{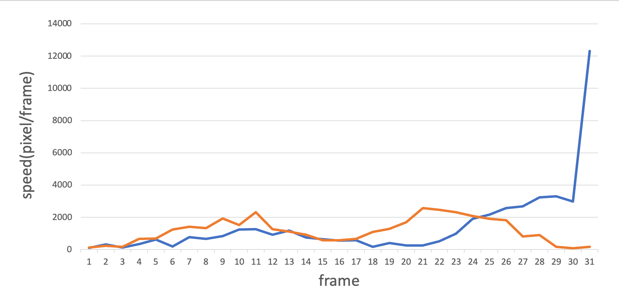}
    \caption{Discrete time-series data of ankle and wrist speeds of pitcher A.}
    \label{fig:pitcherA_speeds}
\end{figure}

\begin{figure}[htbp]
    \centering
    \includegraphics[width=\linewidth]{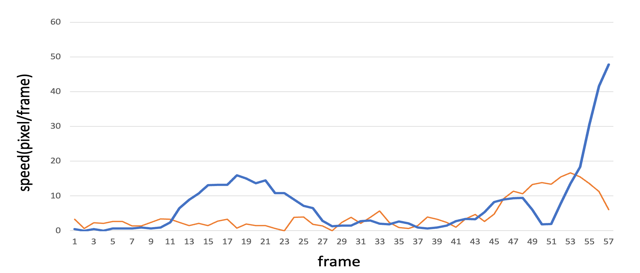}
    \caption{Discrete time-series data of ankle and wrist speeds of pitcher B.}
    \label{fig:pitcherB_speeds}
\end{figure}

Fig.~\ref{fig:pitcherA_speeds} shows the time series data of pitcher A's ankle speed in orange and wrist speed in blue. 
Fig.~\ref{fig:pitcherB_speeds} similarly shows the ankle and wrist speeds of pitcher B.

\subsection{Influence of Camera Angle on Dissimilarity Metric}
In addition, 3D data of the pitching motions of two male subjects, D and E, were obtained using a motion capture system with 12 infrared cameras. 
Both subjects are right-handed.
A total of 16 marker sets were used on the body and the coordinates were obtained by focusing on the ankle of the stepping side (left ankle for the right pitcher) and the wrist of the throwing arm (right wrist for the right pitcher). 
The $+x$ axis is the direction from the pitcher's plate to first base, the $+z$ axis is the direction from home base, and the $+y$ axis is the direction vertically upward. 
The 3D data is used to quantitatively assess the effect of the angle of view of a conventional camera on the dissimilarity assessment. 
When the camera coordinates are obtained by orthogonal projection, the trajectory of each joint in the camera image differs depending on the shooting angle. 
Assuming that the camera position is moved in the horizontal plane by a rotation of $\theta$, the joint coordinates are rotated by $-\theta$ in the horizontal $zx$ plane. 
The camera coordinate $z'$ after rotation is given by~(\ref{eq:z_rotation}),
\begin{equation}
  z'=z \cos(-\theta) - x \sin(-\theta) = z\cos\theta+x\sin\theta.
  \label{eq:z_rotation}
\end{equation}
The time-series of speed is calculated using the 2D coordinates of the horizontal $z'$ and the vertical $y$ after rotation by~(\ref{eq:z_rotation}), and compared with the time series of speed for the reference $\theta=0$. 
With the shooting angle reference $\theta=\SI{0}{\degree}$, 
$$\mathrm{Dis}\left((s_{a0}, s_{w0}),(s_{a\theta}, s_{w\theta})\right)$$ 
is obtained for $\theta = \SI{-80}{\degree}, \SI{-70}{\degree},\dots,+\SI{90}{\degree}$ is obtained by the dissimilarity measure described in the theoretical background. 

To compare the dissimilarity due to individual differences, the time-series of the ankle and wrist at the shooting angle $\theta$ for subject D are $s_{a\theta}^D, s_{w\theta}^D$, respectively, and $s_{a\theta}^E, s_{w\theta}^E$ for subject E, respectively.
We evaluate 
$$\mathrm{Dis}\left((s_{a0}^D, s_{w0}^D),(s_{a\theta}^E, s_{w\theta}^E)\right)$$
$$\mathrm{Dis}\left((s_{a0}^E, s_{w0}^E),(s_{a\theta}^D, s_{w\theta}^D)\right)$$ 
for $\theta = \SI{-80}{\degree}, \SI{-70}{\degree},\dots,+\SI{90}{\degree}$.

\section{Results}
\subsection{Results of Video-Based Body Coordination Analysis}
We evaluate the dissimilarity between the ankle and wrist velocity time series data of the pitching motion of MLB pitchers. 
In this experiment, two videos (A1, A2, B1, B2, C1, C2) were prepared for each of three pitchers (A, B, C). 

\begin{figure}[htbp]
    \centering
    \includegraphics[width=\linewidth]{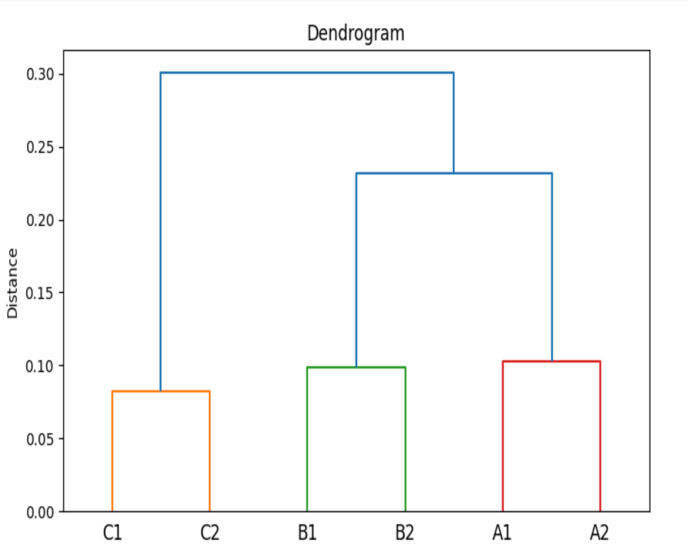}
    \caption{Dendrogram based on the proposed dissimilarity.}
    \label{fig:dendrogram}
\end{figure}

Table~\ref{table:result_videos_3pitchers} summarizes the dissimilarity values among the six analyzed videos.
\begin{table}[htbp]
  \centering
  \caption{Dissimilarity values in 3 pitchers, 2 videos each.}
  \label{table:result_videos_3pitchers}
  \begin{tabular}{c|cc|cc|cc}
     & A1 & A2 & B1 & B2 & C1 & C2\\
    \hline
    A1 & 0 & \underline{0.1027} & 0.2131 & 0.1229 & 0.1842 & 0.2199\\
    A2 & --- & 0 & 0.2227 & 0.1327 & 0.1532 & 0.2811\\
    \hline
    B1 & --- & --- & 0 & \underline{0.0986} & 0.1883 & 0.1911\\
    B2 & --- & --- & --- & 0 & 0.2491 & 0.2003\\
    \hline
    C1 & --- & --- & --- & --- & 0 & \underline{0.0819}\\
    C2 & --- & --- & --- & --- & --- & 0
  \end{tabular}
\end{table}
A dendrogram visualizing the similarity between the video data of each subject (A1, A2, B1, B2, C1, C2) is shown in Fig.~\ref{fig:dendrogram}. 
Ward's method was used as the clustering method and clusters were merged based on the pairwise dissimilarity between the data. 
The horizontal axis shows the label of the data and the vertical axis shows the distance between the clusters, with the closest data being merged at a lower position.

\begin{figure}[htbp]
    \centering
    \includegraphics[width=\linewidth]{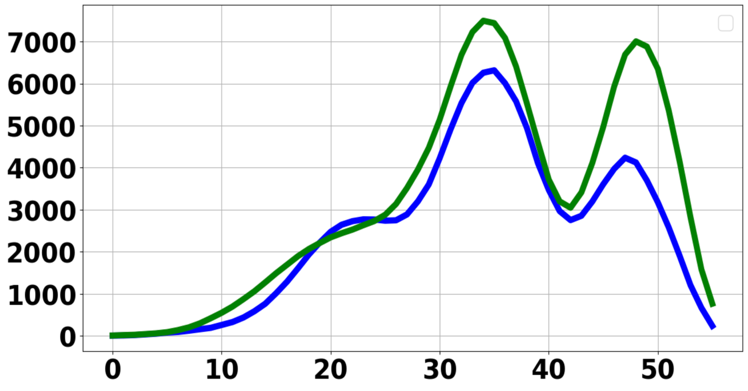}
    \caption{Convolution vectors computed from ankle-wrist velocity signals for two pitching motions by Pitcher A.}
    \label{fig:convolution_A}
\end{figure}

\begin{figure}[htbp]
    \centering
    \includegraphics[width=\linewidth]{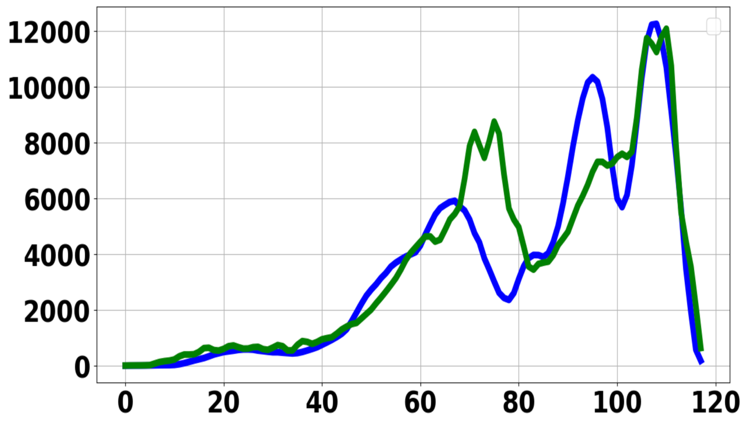}
    \caption{Convolution vectors computed from ankle-wrist velocity signals for two pitching motions by Pitcher B.}
    \label{fig:convolution_B}
\end{figure}

\begin{figure}[htbp]
    \centering
    \includegraphics[width=\linewidth]{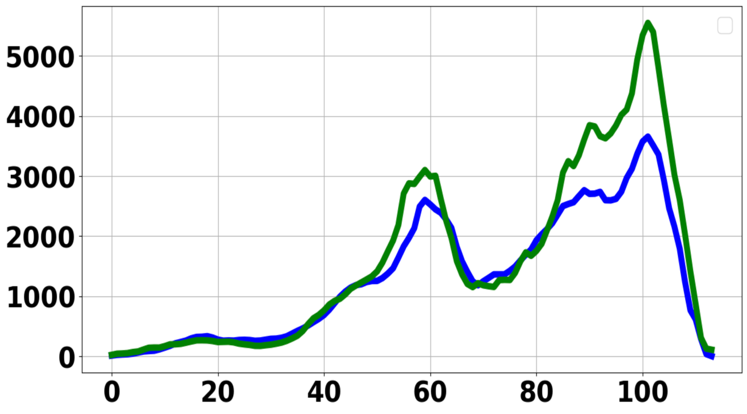}
    \caption{Convolution vectors computed from ankle-wrist velocity signals for two pitching motions by Pitcher C.}
    \label{fig:convolution_C}
\end{figure}

\begin{figure}[htbp]
    \centering
    \includegraphics[width=\linewidth]{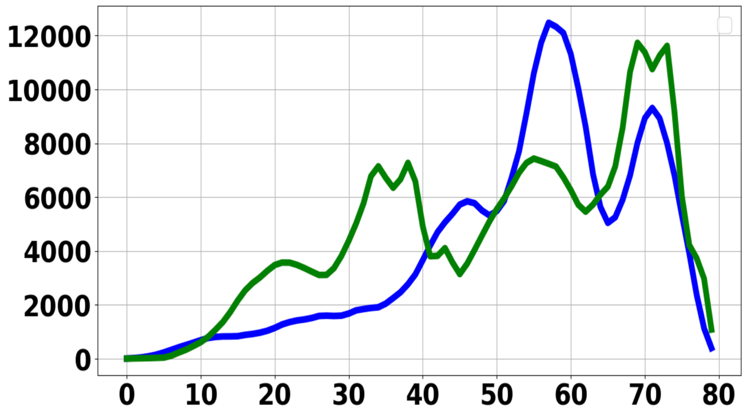}
    \caption{Convolution vectors computed from ankle-wrist velocity signals for pitching motions A1 \& B2.}
    \label{fig:convolution_AandB}
\end{figure}

\begin{figure}[htbp]
    \centering
    \includegraphics[width=\linewidth]{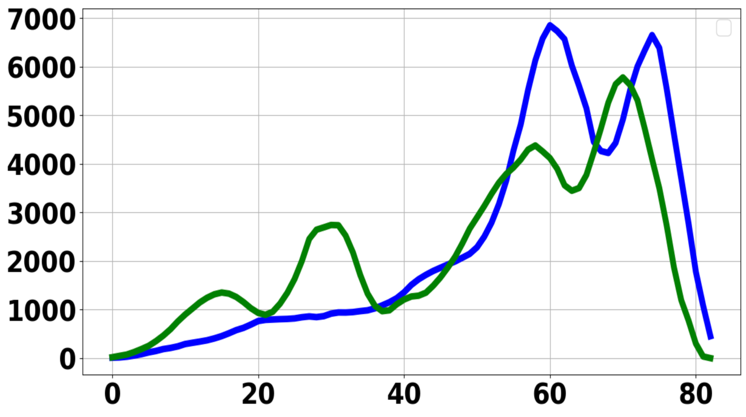}
    \caption{Convolution vectors computed from ankle-wrist velocity signals for pitching motions A2 \& C1.}
    \label{fig:convolution_AandC}
\end{figure}

\begin{figure}[htbp]
    \centering
    \includegraphics[width=\linewidth]{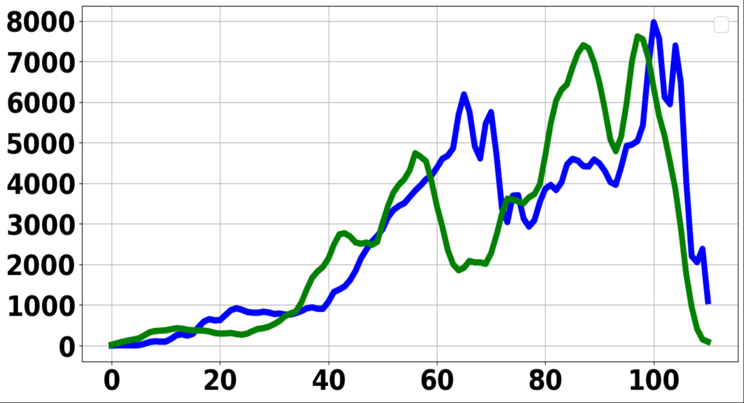}
    \caption{Convolution vectors computed from ankle-wrist velocity signals for pitching motions B1 \& C1.}
    \label{fig:convolution_BandC}
\end{figure}

The $(M+N-1)$-dimensional vectors $u,v$ calculated for the two videos of the same pitcher for the three pitchers are visualized in Figs.~\ref{fig:convolution_A},~\ref{fig:convolution_B} and~\ref{fig:convolution_C}.
Fig.~\ref{fig:convolution_AandB}, \ref{fig:convolution_AandC} and~\ref{fig:convolution_BandC} visualize the $(M+N-1)$-dimensional vectors $u$ and $v$ calculated for the two videos between different pitchers.

\subsection{Results of Influence of Camera Angle on Dissimilarity Metric}
\begin{figure}[htbp]
    \centering
    \includegraphics[width=\linewidth]{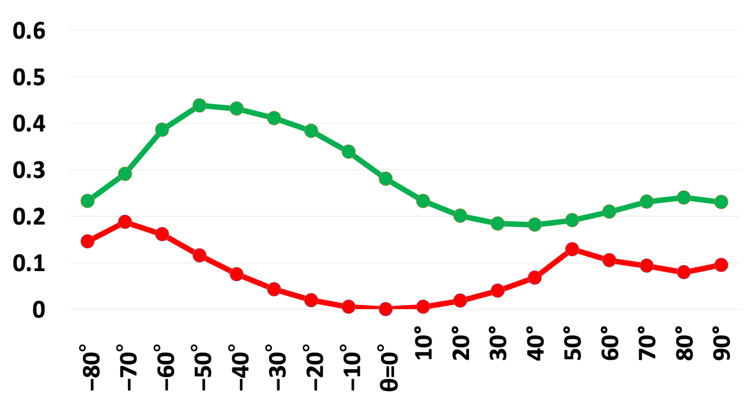}
    \caption{Dissimilarity change with shooting angle. Motion of subject D at $\theta=\SI{0}{\degree}$ is the base.}
    \label{fig:shooting_angle_D0}
\end{figure}

\begin{figure}[htbp]
    \centering
    \includegraphics[width=\linewidth]{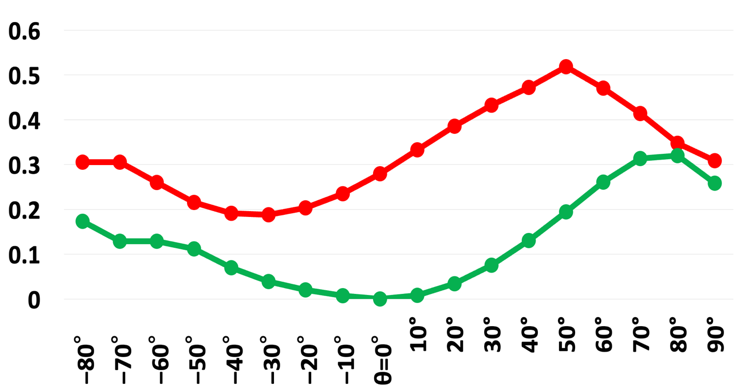}
    \caption{Dissimilarity change with shooting angle. Motion of subject E at $\theta=\SI{0}{\degree}$ is the base.}
    \label{fig:shooting_angle_E0}
\end{figure}

Figs.~\ref{fig:shooting_angle_D0} and \ref{fig:shooting_angle_E0} show the change of dissimilarity for each shooting angle of the two motions by two subjects.
In Fig.~\ref{fig:shooting_angle_D0}, $\mathrm{Dis}\left((s_{a0}^D, s_{w0}^D),(s_{a\theta}^D,s_{w\theta}^D)\right)$ is shown red and $\mathrm{Dis}\left((s_{a0}^D,s_{w0}^D ),(s_{a\theta}^E,s_{w\theta}^E )\right)$ are shown in green for $\theta=\SI{-80}{\degree}, \SI{-70}{\degree}, \dots, +\SI{90}{\degree}$.
In Fig.~\ref{fig:shooting_angle_E0}, $\mathrm{Dis}\left((s_{a0}^E, s_{w0}^E),(s_{a\theta}^D,s_{w\theta}^D)\right)$ is shown red and $\mathrm{Dis}\left((s_{a0}^E, s_{w0}^E ),(s_{a\theta}^E,s_{w\theta}^E )\right)$ are shown in green.

\section{Discussions}
In this study, we proposed a dissimilarity scale to quantitatively assess the body coordination of pitching in a new framework by considering it as a synergistic coordination system. 
We discuss the implications of our results in detail below.

The results of Figs.~\ref{fig:convolution_A},~\ref{fig:convolution_B} and~\ref{fig:convolution_C} show that the computed dissimilarity values between different motions performed by the same pitcher are relatively small. 
In contrast, the results of Figs.~\ref{fig:convolution_AandB},~\ref{fig:convolution_AandC} and~\ref{fig:convolution_BandC} show that the dissimilarity between different pitchers is large. 
Dissimilarity values ranged from 0.08 to 0.10 between motions performed by the same pitcher, whereas they ranged from 0.18 to 0.28 between motions of different pitchers.
This suggests that our metric accurately captures coordination characteristics regardless of moderate variations in horizontal camera angles. 
The dendrogram in Fig.~\ref{fig:dendrogram} shows that the pitching motions of the same pitcher are classified into the same cluster and could be clearly distinguished from those of different pitchers.
This demonstrates that our proposed metric effectively distinguishes individual differences in synergistic coordination. 

From the results of Figs.~\ref{fig:shooting_angle_D0} and~\ref{fig:shooting_angle_E0}, we can see that the dissimilarity caused by the difference in the shooting angles is smaller than the individual differences in the body coordination of the movements performed with the 2D images. 
When $|\theta|<\SI{40}{\degree}$, the dissimilarity with the case of $\theta=\SI{0}{\degree}$ is small for the same movement. 
This suggests that the dissimilarity can be evaluated correctly regardless of the shooting angle in the horizontal range.

However, in this experiment, we assumed an idealized scenario where all joint marker coordinates are clearly visible and free from occlusion.
The joint coordinate estimation accuracy using OpenPose is influenced by occlusion.
Future work should address scenarios involving occlusion and varying confidence levels of joint coordinate estimation using OpenPose, enhancing the metric’s applicability to more realistic conditions.

\section{Conclusion}
In this study, we introduced a convolution-based dissimilarity metric to quantitatively assess dynamic body coordination between pairs of body segments. 
Our metric explicitly captures the gain and phase-lag characteristics of body segment interactions without unnecessary temporal warping. 
Empirical experiments validated the method’s robustness against changes in camera angle and scaling differences, demonstrating enhanced discriminative capability compared to conventional DTW and recent deep learning-based approaches. 
The results highlight the potential for broader applications in fields requiring precise analysis of rhythmic variations and genuine synergistic coordination.


%








\bibliographystyle{IEEEtran}
\bibliography{./refs.bib}
%

%








\end{document}